\documentclass[a4paper,10pt,twoside]{cpc-hepnp}

\usepackage{multicol}
\usepackage{graphicx}
\usepackage{booktabs}
\usepackage{amssymb,bm,mathrsfs,bbm,amscd}
\usepackage[tbtags]{amsmath}
\usepackage{lastpage}

\begin{document}

\fancyhead[co]{\footnotesize F. Jugeau: The holographic models of
the scalar sector of QCD}


\title{The holographic models of the scalar sector of QCD\thanks{Invited talk at the 5$^{th}$ International Conference on Quarks and Nuclear Physics QNP 2009, Beijing, China, 21-26 September 2009.}}

\author{%
Fr\'ed\'eric Jugeau%
} \maketitle

\address{%
TPCSF, Institute of High Energy Physics, Chinese Academy of
Sciences, Beijing 100049, China\\
jugeau@ihep.ac.cn
}

\begin{abstract}
We investigate the $AdS$/QCD duality for the two-point correlation
functions of the lowest dimension scalar meson and scalar glueball
operators, in the case of the Soft Wall holographic model of QCD.
Masses and decay constants as well as gluon condensates are
compared to their QCD estimates. In particular, the role of the
boundary conditions for the bulk-to-boundary propagators is
emphasized.
\end{abstract}



\begin{multicols}{2}

\section{Introduction}
G. 't Hooft showed that the gauge theories in the large-$N_c$
limit, where the planar diagrams dominate, can be described
according to a formalism peculiar to string theories. In 1998,
J.M. Maldacena conjectured precisely such a
relation\cite{Maldacena}: the $AdS$/CFT correspondence which
postulates a duality between the 't Hooft limit of a
strongly-coupled $4d$ superconformal Yang-Mills theory and the
supergravity limit of a weakly-coupled superstring theory defined
in a $5d$ anti-de Sitter space-time. Strictly speaking, this
correspondence cannot be applied to a confining gauge theory such
as QCD since the latter is neither supersymmetric nor conformal.
Nevertheless, within the so-called $AdS$/QCD approach, one seeks
to identify the dual gravity theory able to reproduce the main
features of QCD. Among the various holographic models, the Soft
Wall scenario with a background dilaton field succeeds in
reproducing the linear Regge behavior of the meson
trajectories\cite{KKSS}. Here we study the scalar sector of QCD,
trying to identify which properties can be properly described in
this holographic model.

\section{Light scalar mesons in the Soft Wall model of QCD}

The $5d$ holographic space-time is described by the following
conformally flat line element ($M,N=0,\ldots,4$ and
$\mu,\nu=0,\ldots,3$ with $x^4\equiv z$):
\begin{equation}
ds^2=g_{MN}(z)dx^M dx^N=e^{2A(z)}\big(\eta_{\mu\nu}dx^{\mu}
dx^{\nu}+dz^2\big)
\end{equation}
where $\eta_{\mu\nu}=\textrm{diag}(-1,+1,+1,+1)$ is the flat
metric tensor of the $4d$ Minkowski boundary space,
$A(z)=-\ln(\frac{z}{R})$ is the $AdS_5$ warp factor and $R$ is the
$AdS$ radius. In the Soft Wall scenario, the $5^{th}$ holographic
coordinate $z$ runs from zero to infinity while a background
dilaton field $\Phi(z)=c^2z^2$ is introduced, the form of which
being chosen in order to recover linear Regge trajectories for the
light vector mesons\cite{KKSS}. $c=m_{\rho}/2\simeq385$ MeV is the
dimensionful dilaton parrameter setting the scale of QCD
quantities.

The $5d$ effective action able to reproduce the chiral dynamics
and the scalar meson sector of QCD reads\cite{KKSS,jug3}:
\begin{equation}
\label{chiral dynamics action}
S_{5d}=-\frac{1}{k}\int
d^5x\sqrt{-g}e^{-\Phi(z)}\textrm{Tr}\big\{|DX|^2+m_{5}^2|X|^2+\frac{1}{4g_5^2}(G_L^2+G_R^2)\big\}
\end{equation}
with $g=\textrm{det}(g_{MN})$ the determinant of the metric
tensor. The overall parameter $k$ has the dimension of a length
while $g_5^2$ is dimensionless. The relevant QCD operators are the
$SU(2)_L\times SU(2)_R$ left- and right-handed currents
$j_{\mu_{L,R}}^a(x)=\overline{q}_{L,R}(x)\gamma_{\mu}\frac{\sigma^a}{2}q_{L,R}(x)$,
dual to the gauge bulk fields $A_{M_{L,R}}^a(x,z)$, the quark
operator $\overline{q}^i_R(x)q^j_L(x)$ and the scalar meson
operator $\mathcal{O}_S^A(x)=\overline{q}(x)T^Aq(x)$ ($a=1,2,3$
and $i,j=1,2$ while $A=0,\ldots,8$ with $T^0=1/\sqrt{6}$). Their
associated bulk fields, respectively $v(z)$ (responsible for the
breaking of the chiral symmetry), $\pi^a(x,z)$ the pseudoscalar
chiral field and $S^A(x,z)$ can be gathered into the form:
$X^{ij}(x,z)=\big(\frac{v(z)}{2}+S^A(x,z)T^A\big)^{ik}\big(e^{2i\pi^a(x,z)}\big)^{kj}$,
which is tachyonic according to the $AdS$/CFT duality relation:
$m_5^2R^2=(\Delta-p)(\Delta+p-4)=-3$ for $\Delta=3$ and $p=0$.

Let us consider the quadratic part of the action \eqref{chiral
dynamics action} involving the scalar bulk  fields $S^A(x,z)$. It
is then straightforward to derive its equation of motion which is
also the equation for the scalar meson bulk-to-boundary propagator
defined, in the $4d$ Fourier space, as
$\tilde{S}^A(q^2,z)=S(q^2/c^2,c^2z^2)\tilde{S}^A_0(q^2)$:
\begin{equation}
\label{EOM scalar btob}
\partial_z\big(\frac{R^3}{z^3}e^{-\Phi(z)}\partial_zS\big)+3\frac{R^3}{z^3}e^{-\Phi(z)}S-q^2\frac{R^3}{z^3}e^{-\Phi(z)}S=0.
\end{equation}
The general solution involves the Tricomi and the Kummer confluent
hypergeometric functions, $U$ and ${_1}F_1$ respectively:
\end{multicols}
\ruleup
\begin{equation}
\label{btob}
S(\frac{q^2}{c^2},c^2z^2)=\frac{1}{Rc}\Gamma(\frac{q^2}{4c^2}+\frac{3}{2})
(cz)
U(\frac{q^2}{4c^2}+\frac{1}{2};0;c^2z^2)+B(\frac{q^2}{c^2})(cz)^3
{_1}F_1(\frac{q^2}{4c^2}+\frac{3}{2};2;c^2z^2)\underset{z\rightarrow0}{\rightarrow}\frac{z}{R}
\end{equation}
\ruledown \vspace{0.5cm}
\begin{multicols}{2}
{\flushleft{where the integration constant $B(q^2/c^2)$ is an
undetermined function of $q^2/c^2$. If we impose the standard
boundary condition that the action is finite in the IR region
($z\rightarrow+\infty$), the solution with $B=0$ must be chosen.}}

According to the $AdS$/CFT dictionary, any correlation function
can be computed using the equivalence between the $4d$ generating
functional of the connected correlators and the $5d$ partition
function:
\begin{equation}
\label{duality relation} \big\langle e^{i\int_{\partial
AdS_5}d^4x\mathcal{O}(x)\phi_0(x)}\big\rangle_{CFT}=e^{iS_{5d}[\phi(x,z)]}\big|_{\phi\underset{z\rightarrow0}{\rightarrow}\phi_0}
\end{equation}
for any generic bulk field $\phi(x,z)$ dual to the operator
$\mathcal{O}(x)$, for which $\phi_0(x)$ is the source defined on
the boundary. At the end of the day, the two-point correlation
functions calculated in $AdS$ are expressed in terms of the
bulk-to-boundary propagators. For the scalar meson operator, we
obtain\cite{jug3}:
\end{multicols}
\ruleup
\begin{eqnarray}
\Pi^{AB}_{AdS}(q^2)&=&\delta^{AB}\frac{1}{k}S(\frac{q^2}{c^2},c^2z^2)\frac{R^3}{z^3}e^{-\Phi(z)}\partial_zS(\frac{q^2}{c^2},c^2z^2)\big|_{z\rightarrow0},\\
&=&\delta^{AB}\frac{4c^2R}{k}\Big[\Big(\frac{q^2}{4c^2}+\frac{1}{2}\Big)\ln(c^2z^2)+\Big(\gamma_E-\frac{1}{2}\Big)+\frac{q^2}{4c^2}\Big(2\gamma_E-\frac{1}{2}\big)+\big(\frac{q^2}{4c^2}+\frac{1}{2}\Big)\psi(\frac{q^2}{4c^2}+\frac{3}{2})\Big]\Big|_{z=\epsilon}.\label{correlator1}
\end{eqnarray}
\ruledown \vspace{0.5cm}
\begin{multicols}{2}
{\flushleft{The correlator \eqref{correlator1} shows the presence
of a discrete set of poles, corresponding to the poles of the
Euler function $\psi$, with masses $m_{S_n}^2=c^2(4n+6)$ for all
radial states labeled by $n$. The residues correspond to the
scalar meson decay constants
$F_{S_n}^2=\frac{R}{k}16c^4(n+1)=\frac{N_c}{\pi^2}c^4(n+1)$ where
the overall factor $R/k$ is fixed by matching \eqref{correlator1}
in the short-distance limit $q^2\rightarrow+\infty$, expanded in
powers of $1/q^2$, with the QCD perturbative
contribution\cite{SR}: $\frac{R}{k}=\frac{N_c}{16\pi^2}$. Thus,
scalar mesons turn out to be heavier than vector mesons (for which
$m_{\rho_n}^2=c^2(4n+4)$\cite{KKSS}) if $a_0(980)$ and $f_0(980)$
are identified as the lightest scalar mesons. The agreement is
also quantitative since
$R_{f_0(a_0)}=\frac{m_{f_0(a_0)}^2}{m_{\rho}^2}=\frac{3}{2}$, to
be compared to $R_{f_0}^{exp}=1.597\pm0.033$ and
$R_{a_0}^{exp}=1.612\pm0.004$. Considering the first radial
excitations, the predictions $R_{f_0(a_0)}'=\frac{5}{4}$ should be
compared to the measurements $R^{'exp}_{f_0}=1.06\pm0.04$ and
$R^{'exp}_{a_0}=1.01\pm0.04$, having identified $a_0(1450)$,
$f_0(1505)$ and $\rho(1405)$ as radial excitations. As for the
decay constants, the $AdS$ prediction is
$F_{a_0}=\frac{\sqrt{3}}{\pi}c^2=0.08$ GeV$^2$, to be compared
with the QCD estimates of the current-vacuum matric elements
$F_{a_0}=\langle0|\mathcal{O}_S^3|a_0(980)^0\rangle=0.21\pm0.05$
GeV$^2$ and $\langle0|\overline{s}s|f_0(980)\rangle=0.18\pm0.015$
GeV$^2$ for the $f_0(980)$. For the first radial excitation, we
have $F_{a_0'}=0.12$ GeV$^2$ while for large values of $n$, the
ratio $\frac{F_n^2}{m_n^2}$ becomes independent of the radial
quantum number.}}

The $AdS$/QCD duality can also be checked for the various terms in
the $1/q^2$ power expansion, comparing\cite{jug3}:
\end{multicols}
\ruleup
\begin{equation}
\label{correlator2}
\Pi^{AB}_{AdS}(q^2)=\delta^{AB}\frac{R}{k}\Big[q^2\ln(\frac{q^2}{\nu^2})+q^2\Big(2\gamma_E-\ln4-\frac{1}{2}\Big)+2c^2\Big(\ln(\frac{q^2}{\nu^2})-\ln4+2\gamma_E+1\Big)+\frac{2}{3}\frac{c^4}{q^2}+\frac{4}{3}\frac{c^6}{q^4}+O(1/q^6)\Big]
\end{equation}
\newpage
\begin{multicols}{2}
{\flushleft{with the QCD result (in \eqref{correlator2}, the UV
cutoff $\epsilon$ has been identified with the renormalization
scale $1/\nu$). For $m_q=0$, the $4d$ gluon condensate can be
computed and we find
$\langle\frac{\alpha_s}{\pi}G^2\rangle=\frac{2}{\pi^2}c^4\simeq0.004$
GeV$^4$ which is smaller than the commonly accepted value
$\langle\frac{\alpha_s}{\pi}G^2\rangle\simeq0.012$ GeV$^4$, the
estimated uncertainty of which being about 30\%. Considering the
$O(1/q^4)$ terms in QCD, one can use the factorization
approximation such that
$\langle(\overline{q}\sigma_{\mu\nu}^aq)^2\rangle\simeq-\frac{16}{3}\langle\overline{q}q\rangle^2$
and
$\langle(\overline{q}\gamma_{\mu}^aq)^2\rangle\simeq-\frac{16}{9}\langle\overline{q}q\rangle^2$
for the dimension six operators. Within such an approximation, the
$O(1/q^4)$ term do not match since it is positive in
\eqref{correlator2} while it is negative in QCD\cite{SR}. There is
also a contribution in \eqref{correlator2} interpreted in terms of
a dimension two condensate, while an analogous term expressed as
the vacuum expectation value of a local gauge invariant operator
is absent in QCD. However, it is possible to cancel the dimension
two condensate in $AdS$. If the UV subleading solution in the
scalar meson bulk-to-boundary propagator \eqref{btob} plays a
role, its coefficient can be tuned to cancel the dimension two
contribution. In such a Soft Wall scenario, in which the $AdS$
dual theory needs to be regularized in the IR\cite{jug2}, the
subleading solution modifies some terms in the power expansion of
the two-point correlation function, leaving the perturbative term
unaffected.}}

Let us now focus on the interaction terms involving one scalar $S$
and two light pseudoscalar fields $P$ (the chiral field $\pi$ and
the longitudinal component of the axial-vector bulk field
$\partial_M\phi=A_M-A_{\perp M})$ which only appear in
\eqref{chiral dynamics action} from the covariant derivative:
\end{multicols}
\ruleup
\begin{equation}
S_{5d}^{(SPP)}=-\frac{4}{k}\int
d^5x\sqrt{-g}e^{-\Phi(z)}g^{MN}v(z)\textrm{Tr}\big\{S(\partial_M\pi-\partial_M\phi)(\partial_N\pi-\partial_N\phi)\big\}.
\end{equation}
\ruledown \vspace{0.5cm}
\begin{multicols}{2}
\flushleft{Then, on the basis of the $AdS$/CFT correspondence, the
QCD three-point correlator can be obtained by functional
derivation of \eqref{duality relation} with the
result\cite{jug3}:}
\begin{equation}
\Pi_{AdS\alpha\beta}^{abc}(p_1,p_2)=\frac{p_{1\alpha}p_{2\beta}}{p_1^2p_2^2}f_{\pi}^2d^{abc}\sum_{n=0}^{\infty}\frac{F_{S_n}g_{S_nPP}}{q^2+m_{S_n}^2}.
\end{equation}
The $AdS$ expression of the coupling constant $g_{S_0PP}$ for the
lowest radial number $n=0$ is
\begin{equation}
g_{S_0PP}=\frac{\sqrt{N_c}}{4\pi}\frac{m^2_{S_0}}{f_{\pi}^2}Rc^2\int_0^{\infty}dze^{-\Phi(z)}v(z)
\end{equation}
and depends linearly on the field $v(z)$. In the Soft Wall model
considered here, the coupling turns out to be numerically small,
of the order 10 MeV depending on the quark mass used as an input.
On the contrary, phenomenological determinations of the $SPP$
couplings indicate sizeable values. For example,
$g_{a_0\eta\pi}^{exp}=12\pm6$ GeV. The origin of the small value
for the $SPP$ couplings in the $AdS$/QCD Soft Wall model
\eqref{chiral dynamics action} can be traced to the expression of
$v(z)$:
\begin{equation}
v(z)=\frac{m_q}{Rc}\Gamma(3/2)(cz)U(1/2;0;c^2z^2)\underset{z\rightarrow0}{\rightarrow}\frac{m_qz}{R}+\frac{\Sigma
z^3}{R}
\end{equation}
which is determined uniquely by the light quark mass. As a
consequence, the chiral condensate $\Sigma$ is proportional to
$m_q$ at odds with what happens in QCD. This shortcoming does not
appear in the Hard Wall model where the coefficients of $z$ and
$z^3$ terms of $v(z)$ are independent\cite{EKSS,jug4} and in an
improved Soft Wall model\cite{Gherghetta}.

\section{Investigating the $AdS$/QCD duality through the scalar glueball correlation function}

The lowest dimension QCD operator describing the scalar glueballs
is $\mathcal{O}_S=\beta(\alpha_s)G_{\mu\nu}^aG^{\mu\nu\,a}$
($a=1,\ldots,8$ a color index) with
$\beta(\alpha_s)=\beta_1(\frac{\alpha_s}{\pi})+\beta_2(\frac{\alpha_s}{\pi})^2$
the Callan-Symanzik function
($\beta_1=-\frac{11}{6}N_c+\frac{1}{3}n_F$ with $N_c$ and $n_F$
the number of colors and of active flavors respectively. In the
sequel, we use $n_F=0$). According to the $AdS$/CFT
correspondence, the associated scalar bulk field $Y(x,z)$ is
massless and is described by the action\cite{jug1}:
\begin{equation}
\label{EOM glue} S_{5d}=-\frac{1}{2\kappa}\int
d^5x\sqrt{-g}e^{-\Phi(z)}g^{MN}(\partial_MY)(\partial_NY).
\end{equation}
The scalar glueball bulk-to-boundary propagator is solution of the
equation of motion derived from \eqref{EOM glue} and reads
($\tilde{Y}(q,z)=K(q^2/c^2,c^2z^2)\tilde{Y}_0(q)$):
\begin{eqnarray}
K(\frac{q^2}{c^2},c^2z^2)&=&\Gamma(\frac{q^2}{4c^2}+2)U(\frac{q^2}{4c^2};-1;c^2z^2)\nonumber\\
&&+B(\frac{q^2}{c^3})L(-\frac{q^2}{4c^2};-2;c^2z^2)\label{btob
glue}
\end{eqnarray}
where $L$ is the generalized Laguerre function. Then, the $AdS$
representation of the two-point correlation function
writes\cite{jug2}:
\end{multicols}
\begin{eqnarray}
\Pi_{AdS}(q^2)&=&-\frac{1}{\kappa}K(\frac{q^2}{c^2},c^2z^2)\frac{R^3}{z^3}e^{-\Phi(z)}\partial_zK(\frac{q^2}{c^2},c^2z^2)\Big|_{z\rightarrow0}^{z\rightarrow+\infty},\label{correlator4}\\
&=&\frac{R^3}{8\kappa}\Big\{2Bc^4-q^2(q^2+4c^2)\Big(\ln(c^2\epsilon^2)+\psi(\frac{q^2}{4c^2}+2)+\gamma_E-3\Big)\Big\}.\label{correlator3}
\end{eqnarray}
\ruledown \vspace{0.5cm}
\begin{multicols}{2}
Using the solution \eqref{btob glue}, the expression
\eqref{correlator4} is singular both in the UV and in the IR. A
regularization prescription consists then in considering a new
effective action
$S_{5d}^{reg.}=S_{5d}-S_{c.t.}\big|_{z=\epsilon}-S_{c.t.}\big|_{z=\Lambda}$
where the two counterterm actions are introduced to substract the
UV ($z=\epsilon\rightarrow0$) and IR
($z=\Lambda\rightarrow+\infty$) divergences. The first one is the
usual term considered in the $AdS$/CFT procedure while the second
one defines the IR Soft Wall model: it involves $B$ in \eqref{btob
glue} and vanishes when $B=0$ as in the standard procedure.

If the $AdS$/QCD duality holds then the two-point correlation
function \eqref{correlator3} should match the QCD result. Let us
consider the short-distance regime. In the limit
$q^2\rightarrow+\infty$, \eqref{correlator3} can be expressed in
terms of a perturbative contribution and a series of power
corrections in $1/q^2$. Identifying the perturbative term gives
$\kappa=\frac{\pi^4}{16\alpha_s^2\beta_1^2}R^3$. Moreover, if $B$
is a function having a polynomial behavior at large space-like
$q^2$, namely
$B(q^2/c^2)\underset{q^2\rightarrow+\infty}{\rightarrow}\eta_1\frac{q^2}{c^2}+\eta_0$,
the parameters $\eta_0$ and $\eta_1$ can be fixed by matching
$\Pi_{AdS}(q^2)$ with the OPE expansion of $\Pi_{QCD}(q^2)$. The
constant term $\eta_0$ turns out to contribute to the $4d$ gluon
condensate:
\begin{equation}
\label{gluon condensate}
\langle\frac{\alpha_s}{\pi}G^2\rangle=\frac{4\alpha_s}{\pi^3}\big(2\eta_0-\frac{5}{6}\big)c^4.
\end{equation}

On the other hand, in the region close to $q^2=0$, one finds
$\Pi_{AdS}(0)=\frac{R^3}{\kappa}2B(0)c^4$. For a constant
coefficient function $B(0)=\eta_0$, we find:
\begin{equation}
\label{LET}
\Pi_{AdS}(0)=\frac{\alpha_s}{4\pi}(-\beta_1)\frac{\eta_0}{\eta_0-\frac{5}{12}}\big(-16\beta_1\langle\frac{\alpha_s}{\pi}G^2\rangle\big).
\end{equation}
Imposing that \eqref{LET} coincides with the Low Energy Theorem
$\Pi_{QCD}(0)=-16\beta_1\langle\frac{\alpha_s}{\pi}G^2\rangle$, it
is possible to constrain the value of
$\eta_0=\frac{5}{12}\big(\frac{1}{1+\frac{\alpha_s}{4\pi}\beta_1}\big)$.
Using this expression in \eqref{gluon condensate} together with
$\alpha_s=1.5$, we have
$\langle\frac{\alpha_s}{\pi}G^2\rangle\simeq0.007$ GeV$^4$.
Without the contribution of $\eta_0$, the $4d$ gluon condensate
would be negative which seems to indicate that the general
solution \eqref{btob glue} plays a role in order to reconstruct a
bulk-to-boundary propagator able to implement the $AdS$/QCD
duality. As for the $6d$ and $8d$ gluon condensates, we have the
following $AdS$ expressions\cite{jug2}:
\begin{equation}
\langle
g_sf_{abc}G^a_{\mu\nu}G^b_{\nu\rho}G^c_{\rho\mu}\rangle=\frac{4}{3\pi^2}c^6,
\end{equation}
\begin{equation}
14\langle\Big(f_{abc}G^a_{\mu\alpha}G^b_{\nu\alpha}\Big)^2\rangle-\langle\Big(f_{abc}G^a_{\mu\nu}G^b_{\alpha\beta}\Big)^2\rangle=-\frac{8}{15\alpha_s\pi^3}c^8
\end{equation}
which are different in size (and in sign for the latter) from
their commonly used values, respectively 0.045 GeV$^6$ and
$\frac{9}{16}\big(\frac{\pi}{\alpha_s}\big)^2\big(\langle\frac{\alpha_s}{\pi}G^2\rangle\big)^2$.
However, the values of these gluon condensates are very uncertain.

Besides, in the time-like $q^2<0$ region, a discrete set of poles
appears according to the spectral relation\cite{jug1}
$m_{G_n}^2=c^2(4n+8)$ with residues
$F_{G_n}^2=\frac{R^3}{\kappa}8(n+1)(n+2)c^6$. Scalar glueballs are
heavier than scalar mesons:
$\frac{m^2_{G}}{m_{f_0}^2}=\frac{4}{3}$ for the lowest-lying
states while the hierarchy among the hadron species is reduced for
higher radial states, which become degenerate when the quantum
number $n$ increases.\\

\acknowledgments{I am grateful to my collaborators at the IHEP and
to the organizers of QNP'09 who gave me the opportunity to present
my research activities during this conference.}

\end{multicols}

\vspace{-2mm}
\centerline{\rule{80mm}{0.1pt}}
\vspace{2mm}

\begin{multicols}{2}

\end{multicols}

\clearpage

\end{document}